\documentclass[10pt,a4paper]{article}
\usepackage[latin1]{inputenc}
\usepackage{amsmath}
\usepackage{amsfonts}
\usepackage{amssymb}
\usepackage{graphicx}
\usepackage{verbatim}

\author{Hamid A. Toussi\\
Department of Mathematics \& Computer Science\\
University of Sistan and Baluchestan\\
Zahedan, Iran\\
e-mail: hamid2c@gmail.com
}
\title{tym: Typed Matlab}
\begin{document}

\maketitle
\begin{abstract}
Although, many scientists and engineers use Octave or MATLAB as their preferred programming language, dynamic nature of these languages can lead to slower running-time of programs written in these languages compared to programs written in languages which are not as dynamic, like C, C++ and Fortran. In this work we developed a translator for a new programming language (tym) which tries to address performance issues, common in scientific programs, by adding new constructs to a subset of Octave/MATLAB language. Our translator compiles programs written in tym, to efficient C++ code. 
\end{abstract}

\section{Introduction}
\label{intro}
Many scientists and engineers use Octave or MATLAB as their preferred programming language. However, dynamic nature of these languages can lead to slower running-time of programs written in these languages compared to programs written in languages which are not as dynamic, like C, C++ and Fortran.

Two dynamic features that contributes to performance issues in major ways are:
\begin{enumerate}
\item Dynamic typing: Types of variables can change during run-time and generally they are not known before run-time.
\item Dynamic resizing: Size of arrays and matrices can change during run-time. Pre-allocation of arrays is not mandatory in Octave/MATLAB and whenever a value is assigned to a location that is not within the range of the array indexes, the array is resized to store the new value.
\end{enumerate}

For example, consider the following function in Octave/MATLAB:
\begin{verbatim}
function z = mmt(x, y)
  z=x*y;
end
\end{verbatim}
Type of variables \verb=x= and \verb=y= can be any allowable type. If we are supposed to compile this function statically, we may have to choose the widest possible type for \verb=x= and \verb=y=. This type is an array of complex numbers. The result would be very inefficient code when parameters are of narrower type (e.g. Integers or even arrays of integers). In these cases they are actually wrapped as arrays of complex numbers.
One way to tackle this problem is to use JIT compilers to compile the program or choose the best previously compiled version at run-time. However, this requires run-time overhead and can be really non-trivial. We would like to follow another path in this paper.

We have developed a new language which is similar to Octave/MATLAB in many ways but has a less dynamic nature. In particular, variables must be declared (with their type) before they are defined or used and arrays must be allocated explicitly. We have called this language, tym (Typed MATLAB).\\
Programs written in tym are translated into C++, the generated C++ program uses Octave library and can be called from the Octave \cite{octave} interpreter. To do this, an oct-file should be created by using mkoctfile.\\
Cython \cite{cython} also takes a similar approach, but, it is based on Python which might not be as common as Octave/MATLAB in scientific and engineering communities. OMPC \cite{ompc} is another compiler that translates MATLAB to Python. We have used some of their routines and their grammatical rules in our code.

\section{Overview}
\begin{figure}
\caption{tym compiler architecture}
\label{tymarch}
\begin{center}
\includegraphics[scale=0.30]{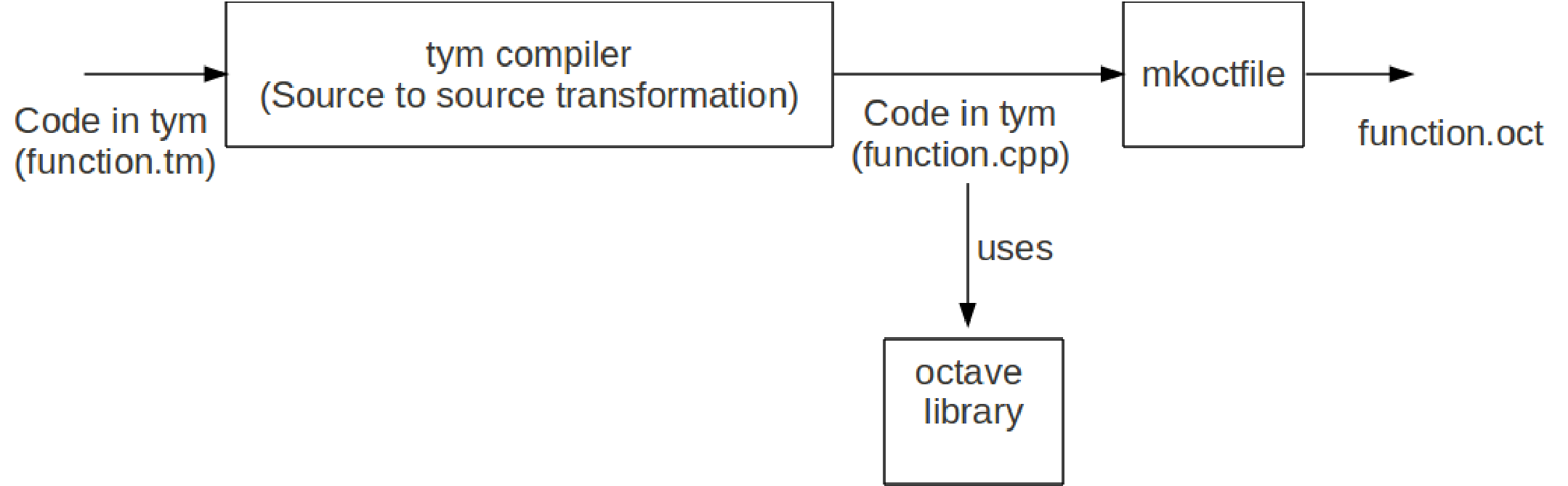}
\end{center}
\end{figure}

You can see different components that are necessary to translate a program in tym to its equivalent module in C++ in Figure~\ref{tymarch}.\\
Currently, a program should be written as a function in tym. Later, this function is transformed to a C++ module and is compiled to an oct-file. The resulted oct-file can be called from the octave interpreter as a function. This has the advantage that the user can change or write parts of her MATLAB program that is computation-intensive as a function in tym then call it from Octave so she has access to all packages in octave-forge (Neural networks, Image processing, Signal processing, ...) while she programs in tym.\\
Resulted C++ module uses Octave library classes and routines to manipulate matrices and other objects in a way that is compatible with Octave. However, since Octave library does not depend on the Octave interpreter and any C++ program can use it independent of the rest of the Octave, it is also possible to convert a program which is written in tym to an executable that can be run without relying on the octave interpreter.

To implement our tym to C++ compiler, we use PLY \cite{ply} which is a Python implementation of compiler construction tools lex and yacc. PLY supports LALR(1) parsing. All grammatical rules should be written in a python module (\verb_tymply.py_ in our compiler). There is a function for every tym programming construct in this module. The grammatical rule which corresponds to a construct is defined in the function's document string. In the body of the function is the action code which is done upon parsing the construct.\\
PLY generates a parser based on the file tymply.py. Since the generated parser is a bottom up LALR(1) parser, we can assume that it does a post-order traversal on Abstract Syntax Tree (AST) and performs some action code upon visiting each AST node. Every AST node is an instance of a tym construct in input text. In our compiler, action codes are responsible for transforming the input program in tym to its equivalent form in C++.\\
AST is an abstract form of parse-tree which is constructed during parsing the input text.

\section{Dynamic Linking and Octave library}
Octave can be dynamically linked with functions which are written in C/C++. Upon linking any of the linked functions can be called from Octave interpreter.\\
Writing Octave compatible functions requires either following the C-Mex interface or using Octave library to manipulate matrices and other objects you would like to pass to Octave. This is necessary to comply to Octave's memory representation of different objects (like matrices). Octave itself uses the Octave library to manipulate matrices and other objects so using it imposes no additional overhead.\\
Using C-Mex interface causes the overhead of converting input parameters of C-Mex routines to its equivalent Octave representation. However, it has the advantage that the functions that uses C-Mex interface can be also linked to MATLAB proprietary software.\\
We have used Octave library so the remainder of this Section is devoted to it and how a C++ function can be linked with Octave.

Consider the example in Figure~\ref{sumsub_cpp}. Using \verb=#include <octave/oct.h>= is necessary so you have access to the required interface to Octava library.\\
\verb=DEFUN_DLD= is a macro that defines the entry point to the function. It has four arguments, name of the function as appear in Octave, list of input parameters to the function, number of output parameters and function's document string.\\
The function should always return an object of type \verb=octave_value_list=. Input parameters are also passed as a \verb=octave_value_list= object (\verb=args=). Input parameters should be extracted from \verb=args= based on their type. Octave passes the input parameters to the function as arrays. However, here, only two integers should be read which are stored in variables \verb=x= and \verb=y= respectively so the first element (indexed by 0) of each input array has been read. Then global variable \verb=error_state= is checked to make sure that input parameters are of expected type. Finally, two values (\verb=a= and \verb=b=) that are supposed to be returned are stored into \verb=retval= which is an \verb=octave_value_list=. Note that operator \verb=()= has been overloaded for \verb=octave_value_list=.\\In order to call this function from Octave, we have to save it as a file that has the same name as the function (\verb=sumsub.cpp= int this example) and run \verb=mkoctfile sumsub.cpp= command. Now we can call this function like any other function from octave interpreter.

\begin{figure}
\caption{A simple function in C++ that can be linked with Octave}
\label{sumsub_cpp}
\begin{footnotesize}
\begin{verbatim}
#include <octave/oct.h>
#include <iostream>
#include <cstdlib>
// File: sumsub.cpp
DEFUN_DLD (sumsub, args, nargout, "do summation and subtraction") {    
    octave_value_list retval;
    if ((args.length()) != 2) { 
        std::cout<<"invalid number of input params\n";
        return retval; 
    }

    int x=args(0).int32_array_value()(0);
    int y=args(1).int32_array_value()(0);
    if (error_state) { 
        std::cout<<"invalid type of input parameters\n";
        return retval;
    }
    int a = x + y;
    int b = x - y;
    retval(0) = a;
    retval(1) = b;
    return retval;
}
\end{verbatim}
\end{footnotesize}
\end{figure}

The base of all matrices and arrays in octave library is \verb=Array= class. All elements of an array is stored linearly in memory (see \verb=data= in Figure~\ref{array_class}). \verb=len= member keeps the number of elements stored in the array. Dimensions and shape of the array can be achieved by referring to \verb=dimensions= vector. Addresses required for doing lookup or assignment operation is calculated based on this vector. Calculated address is an integer that acts as an index for \verb=data= member. The array can be easily reshaped just by making changes to \verb=dimensions= vector.\\In fact, every \verb=Array= object has a pointer to a \verb=ArrayRep= object (\verb=rep=) that contains members like \verb=data= and \verb=len=. This object also keep the number of \verb=Array= objects that share it in \verb=count=. Every Array increments \verb=count= of its \verb=rep= on construction and decrements it on destruction. In \verb=Array='s destructor it is checked that whether \verb=rep->count= has reached zero and free \verb=rep= when it is so.\\
\verb=Array= is a parametrized type which uses parameter \verb=T= to refer to type of the elements stored in the array. Array of different types can be instantiated by passing different types as \verb=T=. 

\begin{figure}
\caption{Definition of Array class in Octave library}
\label{array_class}
\begin{footnotesize}
\begin{verbatim}
template <class T> class Array
{
  class ArrayRep {
    T *data;
    octave_idx_type len;
    int count; // reference count
    ...
  }
  dim_vector dimensions;
  ArrayRep *rep;
  ...
}

\end{verbatim}
\end{footnotesize}
\end{figure}

\verb=Array= class has methods and other data members which are not shown in Figure~\ref{array_class}. Many of its methods like \verb=resize= and \verb=reshape= should be familiar to any Octave/MATLAB user. Array lookup and assignment is done by calling either the methods \verb=xelem= or \verb=checkelem=. The former does not perform bound checking and the latter does.\\
In Octave library \verb=idx_vector= type is used to represent single indices, range slices and colons. Array slicing is done through two family of methods:
\begin{itemize}
\item \verb=index= methods: Whenever, a sliced array is used in an expression, one of these methods will be called. For example, \verb=a(1:4, 2:6)= can be implemented as \verb=a.index(idx_vector(0, 4), idx_vector(1, 6))=. Note that indexing in Octave library is zero based and upper bounds in slices are exclusive.
\item \verb=assign= methods: Indexed assignment to arrays is done by using one of these methods. For example, \verb_a(1:3, :)=b_ can be implemented as \verb=a.assign(idx_vector(0, 3), idx_vector::colon, b)=.
\end{itemize}

\section{Types and Symbol Table Management}
We have used a stack of symbol tables to implement nested scoping. These symbol tables are necessary to keep track of every identifier's type in input program. Other information like the line number where a variable is declared and the line number where it is defined are also kept to do various checks like whether a variable that is used, is declared and defined before its use.\\
Type information is used for various purposes including resolving ambiguity in the grammar in certain cases and very limited type inference and type checking.

\section{Programming and experimenting with tym}
Currently, you can write a function in tym, translate it to C++ and call it from Octave.\\
In contrast to MATLAB/Octave, variables must be declared before they are used. As I write this paper, there are only five types for variables in tym, namely \verb_int_, \verb_real_, \verb_intArray_ and \verb_realArray_. Types \verb_real_ and \verb_float_ correspond to \verb_double_ and \verb_float_ types in C++. In Octave library, arrays of \verb_double_ (i.e \verb_Array<double>_) are used to represent arrays of floating point numbers so only \verb_realArray_ is available in tym to avoid any extra conversion and copying. Hopefully, support for complex numbers and arrays will be added in the future.

There are also some directives in tym language that tells the tym compiler whether to generate code to do array bound checking, initialization of variables and similar stuff. Whenever they come in the tym program they enable/disable the desired feature from the line afterward.\\
As of this writing three direcitves are available. \verb= $ 'zero_based_arrays'= tells tym compiler that matrices are zero based. \verb= $ 'no_init_vars'= tells the compiler not to generate code for initializing variables and \verb= $ 'no_check_ranges'= make the compiler generate code that does not do bound checking. All these directives are off by default. That is,  when no directive has been presented in input program, the compiler generates code for one-based matrices, initializing variables and bound checking which is more consistent with Octave/MATLAB behavior. However, using any of these directive would have positive effect on performance of the generated C++ program, specially \verb= $ 'no_check_ranges'=.

As an example consider the program in Figure~\ref{mult_tym}. This program is a function in tym that multiplies two arrays of type real. All the mentioned directives are used to make it as efficient as possible.

\begin{figure}
\caption{A function in tym that multiplies its parameters}
\label{mult_tym}
\begin{footnotesize}
\begin{verbatim}
$ 'zero_based_arrays'
$ 'no_init_vars'
$ 'no_check_ranges'
% File: mymult.tm
function intArray z = mymult(realArray x, realArray y)
  int d1x = rows(x)
  int d2x = columns(x)
  int d1y = rows(y)
  int d2y = columns(y)

  if (d2x ~= d1y)
    error('incompatible dimensions')
  end

  createArray(z, d1x, d2y)
	
  int i
  int j
  int k
  for i=0:d1x-1
    for j=0:d2y-1
      z(i, j) = 0
      for k=0:d1y-1
        z(i, j) = z(i, j) + x(i, k)*y(k, j)
      end
    end
  end

end

\end{verbatim}
\end{footnotesize}
\end{figure}

The tym compiler would translate the function in Figure~\ref{mult_tym} into the C++ code which is shown in Figure~\ref{mult_cxx}. To call this function from Octave you have to make a dynamically loadable Octave module out of it. To do this execute \verb=mkoctfile mymult.cpp= in a terminal. The mentioned C++ program can be generated by invoking \verb=python tymc.py mymult.tm=.

\begin{figure}
\caption{Translated version of mymult}
\label{mult_cxx}
\begin{footnotesize}
\begin{verbatim}
#include <octave/oct.h>
#include <iostream>
#include <cstdlib>
DEFUN_DLD (mymult, args, nargout, "") {    
    octave_value_list retval;

    NDArray x=args(0).array_value();
    NDArray y=args(1).array_value();
    int d1x = x.rows();
    int d2x = x.columns();
    int d1y = y.rows();
    int d2y = y.columns();
    if ((d2x != d1y)) {
        std::cout<<"error"<<"incompatible dimensions"<<"\n";return retval;
    }
    int32NDArray z(dim_vector( d1x,  d2y));
    int i;
    int j;
    int k;
    for (i = (0); i <= (d1x - 1); i += (1)) {
        for (j = (0); j <= (d2y - 1); j += (1)) {
            z.xelem(i, j) = 0;
            for (k = (0); k <= (d1y - 1); k += (1)) {
                z.xelem(i, j) = 
                   z.xelem(i, j) + x.xelem(i, k) * y.xelem(k, j);
            }
        }
    }
    retval(0) = z;
    return retval;
}

\end{verbatim}
\end{footnotesize}
\end{figure}

As an example for slicing see the program in Figure~\ref{addslice_tym} which would be translated to the C++ code shown in Figure~\ref{addslice_cpp} by the tym compiler. 

\begin{figure}
\caption{A function that add two array slices in tym}
\label{addslice_tym}
\begin{footnotesize}
\begin{verbatim}
$ 'no_check_ranges'
% File: addslice.tm
function intArray z = addslice(intArray x, intArray y)
    if (rows(x) < 3 || columns(x) < 3 || rows(y) < 3 || columns(y) < 3)
       error('Matrices should be of size at least 3x3')
    end
    createArray(z, 3, 3)
    z = x(1:2, 1:2) + y(2:3, 2:3)	
end
\end{verbatim}
\end{footnotesize}
\end{figure}

\begin{figure}
\end{figure}

\begin{figure}
\caption{Translated version of addslice}
\label{addslice_cpp}
\begin{footnotesize}
\begin{verbatim}
#include <octave/oct.h>
#include <iostream>
#include <cstdlib>
DEFUN_DLD (addslice, args, nargout, "") {    
    octave_value_list retval;

    int32NDArray x=args(0).int32_array_value();
    int32NDArray y=args(1).int32_array_value();
    if ((x.rows() < 3 || x.columns() < 3 || y.rows() < 3 || y.columns() < 3)) {
        error("Matrices should be of size at least 3x3");
        return retval;
    }
    int32NDArray z(dim_vector( 3,  3));
    z = ((int32NDArray)x.index(idx_vector(1-1, 2-1+1, 1), idx_vector(1-1, 2-1+1, 1)))
            + ((int32NDArray)y.index(idx_vector(2-1, 3-1+1, 1), idx_vector(2-1, 3-1+1, 1)));
    retval(0) = z;
    return retval;
}
\end{verbatim}
\end{footnotesize}
\end{figure}

You can also try tymc interactively:\\
Type python and then \verb=import tymply as t= in a terminal.
Now you can translate a tym statement to C++ and see the result instantly.\\
Just type \verb=t.yacc.parse("tym_statement")= where \verb=tym_statement= could be any valid tym statement, and press Enter.
\section{Evaluation}
We have done done a limited evaluation using different versions of array multiplication. Results of this evaluation is shown in Table~\ref{times}.\\
Three versions of multiplication have been implemented in tym. \verb=mult-real= is the implementation which is shown in Figure~\ref{mult_tym}. \verb=mult-int= uses \verb=intArray= instead of \verb=realArray=. Version \verb=mult-int-check= is similar to \verb=intArray= except that it does not contain directives for not doing bound checking and variable initialization. Version \verb=mult-octave= is an implementation of array multiplication in Octave/MATLAB. First two arrays of size $100 \times 100$ are filled with random numbers and every version is called to do the multiplication. Then two arrays of size $300 \times 300$ are filled with random numbers and the same process is repeated. Results are shown in Table~\ref{times}. As you can see tym versions are far more efficient than Octave/MATLAB version. This is because of dynamic nature of Octave/MATLAB as explained in Section~\ref{intro}.\\
Among versions which are implemented in tym, \verb=mult-int-check= is the slowest one. The main reason is the bound checking that it does for every array look-up and array assignment.\\
It might be expected that \verb=mult-int= be faster than \verb=mult-real= since operations on integers are faster than operations on floating-points. However, \verb=realArray=s are translated into \verb=NDArray=s which are arrays of doubles (i.e \verb=Array<double>=) but \verb=intArray=s are translated into \verb=int32NDArray=s which are arrays of \verb=octave_int32=. \verb=octave_int32= is a wrapper for integer that is provided by Octave to comply with MATLAB's saturation semantic. Several operators are overloaded for \verb=octave_int32= but to achieve efficiency tymc generates code that does not use these overloaded operators. For example \verb_z(i, j) = z(i, j) + x(i, k)*y(k, j)_ would be translated into
\begin{verbatim}
z(i, j) = z(i, j).value() + x(i, k).value() * y(k, j).value()
\end{verbatim}
in case that \verb_x_ and \verb_y_ and \verb_z_ are \verb_intArray_s. In this way, multiplication and addition are done on plain integers (and not wrapped \verb=octave_int32=s) but this also incurs an additional overhead of calling \verb=value= method of \verb=octave_int32= objects whenever it is used in an expression. That might be the major reason for \verb=mult-int= being slower than \verb=mult-real=.

\begin{table}
\caption{Running times of different versions of multiplication.}
\label{times}
\begin{center}
\begin{tabular} {|c|c|c| }
\hline
\textbf{benchmark} & \textbf{$100 \times 100$} & \textbf{$300 \times 300$}   \\ \hline

mult-int & 0.00457 & 0.1063 \\ \hline
mult-int-check  & 0.0622  & 1.579 \\ \hline
mult-real & 0.002945 & 0.06881 \\ \hline
mult-octave & 19.68  & 155.3 \\ \hline
\end{tabular}
\end{center}
\end{table}

\section*{Acknowledgement}
We also have to credit another work ompc \cite{ompc} which is a MATLAB to python compiler. We have used many of their grammar rules and routines in our work.

\bibliographystyle{plain}
\bibliography{tym-paper.bib}

\end{document}